# Preventing crash in stock market:

# The role of economic policy uncertainty during COVID-19

Peng-Fei Dai[1], Xiong Xiong[1,3], Zhifeng Liu[2] [*], Toan Luu Duc Huynh[4,5], Jianjun Sun[6]

1. College of Management and Economics, Tianjin University, Tianjin, China
2. Management School, Hainan University, Haikou, China
3. China Center for Social Computing and Analytics, Tianjin University, Tianjin, China
4. Chair of Behavioral Finance, WHU – Otto Beisheim School of Management, Vallendar, Germany
5. School of Banking, University of Economics Ho Chi Minh City, Vietnam
6. School of Economics, Hainan University, Haikou, China

**ABSTRACT**

This paper investigates the impact of economic policy uncertainty (EPU) on the crash risk of US stock market during the COVID-19 pandemic. To this end, we use the GARCH-S (GARCH with skewness) model to estimate daily skewness as a proxy for the stock market crash risk. The empirical results show the significantly negative correlation between EPU and stock market crash risk, indicating the aggravation of EPU increase the crash risk. Moreover, the negative correlation gets stronger after the global COVID-19 outbreak, which shows the crash risk of the US stock market will be more affected by EPU during the pandemic.

## 1. Introduction

The economic downturn during the COVID-19 pandemic has resulted in a significant decline in the stock market. Some of the previous studies have examined the impacts of COVID-19 pandemic on the downside risks of stock market. One of the main conclusions is that, in general, the occurrence of COVID-19 causes a direct and significant drop in stock prices (Baker et al. 2020; Al-Awadhi et al. 2020; Ramelli and Wagner 2020; Zhang et al. 2020; Liu et al. 2020). Djurovic et al. (2020) points the Dow Jones Industrial Index has been dropping by 36.4% between February 18, 2020 and March 23, 2020. While from the perspective of volatility risk, COVID-19 will significantly increase the volatility of the stock market (Baek et al. 2020; Onali 2020; Papadamou et al. 2020). Different from the existing literature, this paper focuses on stock market crash risk. Mazur et al. (2020) and Ziemba (2020) study the US stock market crash during the Covid-19 period. However, they treat the crash risk as the extreme downside volatility. In our paper, following Chen et al. (2001) and Kräussl et al. (2016), we measure the crash risk by using the conditional skewness of the equity market return. It is a better way to simultaneously capture the asymmetry and negative extremes of the crash risk (Kim et al. 2011; Kim et al. 2011b, Wen et al. 2019). And then, we conduct an empirical analysis to investigate the impacts of the economic policy uncertainty on the crash risk of the US stock market.

In fact, in the early days of the pandemic, the US stock market has experienced a plunge. From Figure 1, we can see that the stock market has undergone a severe impact from COVID-19. The S&P 500 plummets by one-third in a short period, from 3380 points on February 14, 2020 to 2237 points on March 23, 2020. Our subsequent empirical findings further confirm this intuitive conclusion that COVID-19 negatively affects stock market crash risk. We find that the severity of the pandemic, whose proxy is the growth rate of the daily new confirmed cases, does have a significant negative impact on the conditional skewness of the market return, i.e., the crash risk of stock market. It is also consistent with Liu et al. (2020)'s work on the Chinese equity market, which indicates that the pandemic increases the crash risk of stock market.

However, we have noticed that although the number of daily confirmed cases in the United States continued to rise in the following period, stock prices gradually returned to the level before the pandemic, and even hit a new high in the past three years. It indicates that the severity of the pandemic alone is not enough to explain the stock crash.

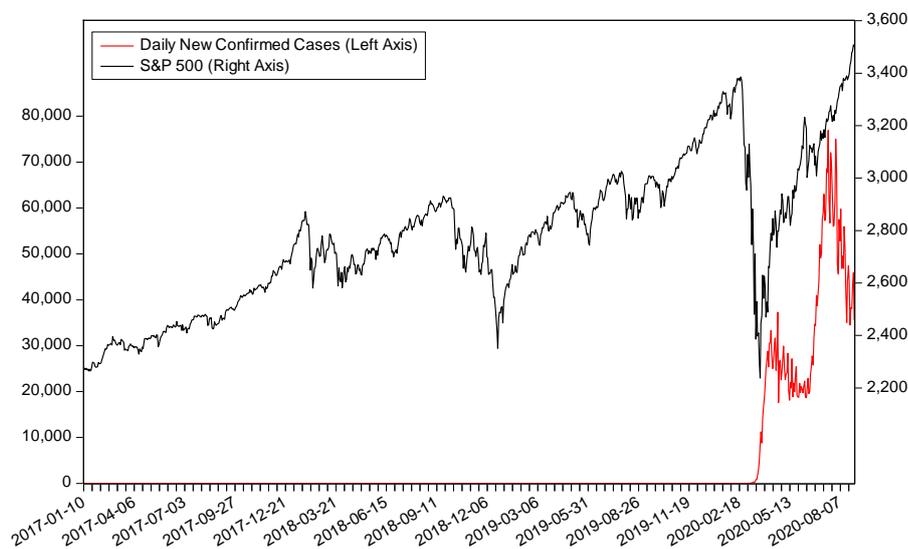

**Figure 1. Daily new confirmed cases and S&P 500 Index**

We argue that one of the reasons for this situation is the reduction of uncertainty in economic policies, which is believed to have a wide-ranging impact on economic and financial activities (Gulen and Ion 2016; Brogaard and Detzel 2015; Dai et al. 2020; Wen et al. 2019; Yousaf and Ali 2020). Figure 2 illustrates the evolving curves of uncertainty indices and S&P 500. As we can see from Figure 2, during the pandemic, economic policy uncertainty (EPU) and stock prices display the opposite trend. In the early stages of the pandemic, as the uncertainty caused by COVID-19 rose sharply, stock prices have also suffered a crash. However, with the successive government measures designed to deal with the pandemic, economic policy uncertainty has gradually decreased, which is what we believe a fundamental reason for the stock market rebound. In other words, we believe that the reduction of economic policy uncertainty during the pandemic will help reduce the crash risk of stock market. Our findings also support this hypothesis. We find that the conditional skewness reacts negatively to the change rate of economic policy uncertainty, indicating that the reduction of economic policy uncertainty can effectively reduce the crash risk of stock market. We further find that this effect only exists during the pandemic, and it is not significant during regular periods. It may mean that the stability of economic policies plays a more critical role in reducing the extremely negative impact of major crisis events.

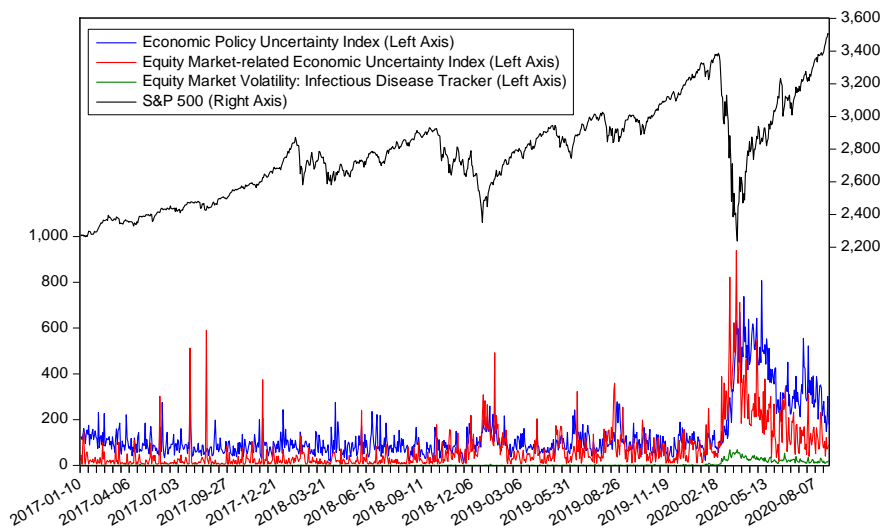

**Figure 2. Economic Policy Uncertainty and the S&P 500 Index**

Before explaining our main contributions, this paper objectives mainly focused on two main folds. First, the estimations of market crash would draw an attention to not only investors but also policymakers to understand how the markets move during this difficult time. Second, the followed regressions with predictive factors will also be our main aims and scopes. Therefore, this study is dedicated to examine the predictive factors of health situation (proxied by number of cases as well as deaths) and the Economic Policy Uncertainties. In doing so, this paper contributes to the extant literature in several ways. First, unlike existing research related to COVID-19, which mainly involves stock market returns and volatility, we focus our attention on the stock market crash risk. Return and volatility are the first and second moments of return distributions, respectively, and we use conditional skewness to measure the crash risk, thus paying attention to the third moment of stock market returns. Second, we confirm the conclusion that COVID-19 will increase the crash risk in the US stock market, which is initially drawn by Liu et al. (2020) in the Chinese stock market. We further find that the increase in the number of confirmed cases can only explain the worsening crash risk of stock market in the early stages of the pandemic. However, it cannot explain why in the middle and late stages, when the number of cases continues increasing, the crash risk will decrease instead. Therefore, the third contribution of our study is that we find that the stability of economic policies can provide a specific explanation for this puzzle. Our findings show that high economic policy uncertainty will increase the crash risk of the US stock market. In contrast, low uncertainty of economic policies can help reduce the likelihood of stock market crashes, especially during the extreme crisis of the COVID-19 global pandemic. More importantly, this paper has important policy implications. Just as our findings show the importance of economic policy stability in reducing the crash risk, we suggest that when facing major crisis events, policymakers should introduce event-response policies as soon as possible. It is helpful to reduce the adverse effects of economic policy uncertainty.

The rest of the paper is organized as follows. Section 2 provides a brief literature review regarding COVID-19 impacts on financial markets. Section 3 introduces the data and methodology. Section 4 shows the empirical results. Conclusions appear in Section 5.

**2. Brief literature review regarding COVID-19 impacts on financial markets**

This section will acknowledges the current literature of COVID-19 pandemic and its impacts on the financial markets. When it comes to the comparison between COVID-19 event and other public health crises, Schell et al. (2020) indicated that the coronavirus outbreak exhibits the significant negative abnormal returns across the majority of equity markets while this phenomenon does not exist in the remaining events such as Ebola, Zika virus, and so forth. In the same vein, the study of Ambros et al. (2020) investigates the role of news on the stock markets returns and volatility. Although this paper sheds a new light on null results of potential channel between returns and pandemic news, the aforementioned paper provides an empirical evidence about the role of number of disease news significantly the European market volatility. This extant literature shapes our motivation to use the GARCH-S, capturing the skewness effects, to measure the possibility of market crashes during this unprecendented events. Instead of using the volatility index, proposed by Alizadeh et al. (2002), this paper would extend the the works of Chen et al. (2001) and Kräusslet al (2016) proxied by the conditional skewness for crashes on the onset of COVID-19 pandemic.

The novel indicator in terms of uncertainties regarding pandemic news, proposed by Baker et al. (2020), effectively explains the market returns and volatility in the United States. The followed studies is ongoing to contribute the new index by aggregating the wide range of economic uncertainties (Altig et al., 2020). Accordingly, the implied volatility rose considerably in February (2020) and fell gradually in the following months since the stock market recovered afterwards. Thus, these papers put the fundamental agruments how the uncertainties are associated with the the stock market jumps. Concomitantly, Wang et al. (2020) explained the efficiency of predictive power of both VIX and EPU on forecasting the equity markets during COVID-19. Interestingly, VIX contains the strongest predictive ability by using the different models. In contrast, the strand of literature also confirmed the gravity of Economic Policy Uncertainty (EPU) as risk factor to predict economic losses (Al-Thaqeb et al., 2020; Youssef et al., 2021; Caggiano et al., 2020; Megaritis et al., 2021).

After reviewing the current literature, we found that the questions regarding the impacts of COVID-19 on the market crashes are still unanswered. Therefore, this paper contributes the ongoing discussion by three main folds. First, the advanced model of GARCH-S benefits the robust rather than using conventional approaches. Additionally, the persistent and autocorrelated skewness would help improve the model performance. Second, using both indicators, including the pandemic situation and Economic Policy Uncertainties, would provide insights about not only the health crisis management but also the market sentiment. Lastly, the studies of advanced economies broaden the current view how these financial markets reacted and jumps (Goodell and Huynh, 2020). However, the emerging markets are still a potential avenue for understanding. Hence, this paper sheds a new light on the Chinese markets, where has the fast resiliency after the economic shocks (Liu et al., 2020).

## 3. Data and methodology

### 3.1. COVID-19 related variables

To measure the severity of the COVID-19 pandemic, we use the logarithmic growth rate of daily confirmed cases (*rCases*) in the US as the proxy. The initial dataset is from *Our World in Data* (https://ourworldindata.org/coronavirus). We also set a dummy variable, *D_epid*, to divide the whole sample into two periods: before- and after- the pandemic. We set the value of this dummy is one after January 21, 2020, the date when the first case of COVID-19 in the United States is confirmed, and zero otherwise.

### 3.2. Economic policy uncertainty indices

We use three economic policy uncertainty related indices based on daily newspaper coverage in the United States in our paper, all of which are proposed by Baker et al. (2016) or Baker et al. (2020). The first variable is the Economic Policy Uncertainty Index for United States (*EPU*), and the second is the Equity Market-related Economic Uncertainty Index (*EMU*). The third is a proxy for COVID-induced economic uncertainty, say, Equity Market Volatility: Infectious Disease Tracker (*EMV-ID*). We take the first variable as our main proxy of Economic Policy Uncertainty, and the other two are used in our robustness checks. In our empirical analysis, we use the logarithmic change rate of EPU indices, so that we get three change rates corresponding to the above indices, namely *rEPU*, *rEMU*, *rEMV-ID*, respectively. If the index has a zero value, we use *log(Index+1)* when calculating the logarithm. All these data are retrieved from FRED, Federal Reserve Bank of St. Louis (https://fred.stlouisfed.org/).

### 3.3. Measuring US stock market crash risk

In our paper, following the works of Chen et al. (2001) and Kräussl et al (2016), the crash risk of the US stock market is proxied by the conditional skewness (*Skew*), which is estimated from the GARCH-S (GARCH with skewness) model. The original idea of using skewness to measure crash risk was put forward by Chen et al. (2001). However, they only calculate a half-a-year horizon skewness from the daily data. To model the daily Euro crash risk, Kräussl et al. (2016) makes use of the Gram-Charlier series expansion method to estimate the conditional skewness. Our GARCH-S model is specified as follows:

$$\begin{aligned}
r_t &= cr_{t-1} + \varepsilon_t; \quad \varepsilon_t \sim (0, \sigma_\varepsilon^2) \\
\varepsilon_t &= h_t^{1/2} \eta_t; \quad \eta_t \sim (0,1); \quad \varepsilon_t | I_{t-1} \sim (0, h_t) \\
h_t &= \alpha_0 + \alpha_1 \varepsilon_{t-1}^2 + \alpha_2 h_{t-1} \\
s_t &= \beta_0 + \beta_1 \eta_{t-1}^3 + \beta_2 s_{t-1}
\end{aligned} \quad (1)$$

Where $r_t$ is the logarithmic return of the S&P 500 index, retrieved from Yahoo Finance (https://finance.yahoo.com/), and $h_t$ is a classical GARCH(1,1) structure[1]. There are two residual forms in the model: the residual $\varepsilon_t$ and the standardized residual $\eta_t$. $I_{t-1}$ represents the information set at the time $t$. $s_t$ is the conditional skewness process and it is the key part in our model. In addition to the constant term, it consists of two parts: the autoregressive part and the lagged return shocks part. Both León et al. (2005) and Kräussl et al. (2016) use the Gram-Charlier series expansion to estimate the model. We will follow their works, and the difference is we truncate at the third moment.

Table 1 reports the estimation results of the GARCH-S model and the standard GARCH model. As can be seen from Table 1, the estimations from GARCH-S model are highly consistent with that from GARCH (1,1) model, which indicates that our model is robust. For comparison purposes, we calculate the estimation of GARCH (1,1) using the Gram-Charlier sequence expansion as well. Comparing the maximum likelihood values of the two models, as well as the AIC, SC and HQ values, we find GARCH-S model is much better than GARCH(1,1). In general, most of the coefficients are significant, suggesting that the model can fit the data well. We focus on the conditional skewness process. As expected, the coefficient of the shock to skewness is positive and significant (0.0361 with a z-statistic 7.2590) and the coefficient of lagged skewness is also positive and significant (0.1544 with a z-statistic 5027.295). The structure of the conditional skewness is very similar to that in the variance case, indicating that the skewness is autocorrelated and persistent.

---

[1] The F-statistics of ARCH test is 104.0580, which shows the GARCH structure is applicable here.

### 3.4. Descriptive statistics

All the data in our paper is daily, and the sample period is from January 2017 to August 2020. Table 2 shows the descriptive statistics of the variables we used, and all variables are stationary.

**Table 2**
Descriptive statistics

| Variables | Mean | Min | Max | Std. Dev. | ADF |
| --- | --- | --- | --- | --- | --- |
| r | 0.0005 | -0.1277 | 0.0897 | 0.0131 | -8.7374*** |
| Skew | 0.0005 | −0.3257 | 0.2094 | 0.0567 | −24.6753*** |
| rCases | 0.0114 | −2.9444 | 2.9444 | 0.2256 | −5.9381*** |
| rEPU | 0.0004 | −1.7103 | 2.3038 | 0.4840 | −24.6728*** |
| rEMU | 0.0007 | −4.1866 | 4.2360 | 1.0065 | −22.0578*** |
| rEMV_ID | 0.0034 | −1.9782 | 1.9782 | 0.4591 | −20.6820*** |

**Notes:** ***, **, * represent statistical significance at 1%, 5%, and 10% levels, respectively. The total observations are 917.

There are several worthy noting points from the descriptive statistics of our variables. First, our variables are stationary in the original level, implying the validity to be employed in the time series models to avoid the spurious results. In addition, the

**Table 1.**

Estimation results of GARCH-S model and GARCH (1,1) model

| GARCH-S model | | | | GARCH (1,1) model | | | |
|---|---|---|---|---|---|---|---|
| Parameter | Value | Parameter | Value | Parameter | Value | Parameter | Value |
| $\mu$ | −0.0425*** | $\beta_0$ | 0.0000 | $c$ | -0.0456 | $\beta_0$ | N/A |
| | (-63.90) | | (0.78) | | (-1.1895) | | |
| $\alpha_0$ | 0.0000*** | $\beta_1$ | 0.0361*** | $\alpha_0$ | 0.0000*** | $\beta_1$ | N/A |
| | (62.31) | | (7.2590) | | (6.5838) | | |
| $\alpha_1$ | 0.2065*** | $\beta_2$ | 0.1544*** | $\alpha_1$ | 0.2292*** | $\beta_2$ | N/A |
| | (88.01) | | (5027.2950) | | (7.9513) | | |
| $\alpha_2$ | 0.7720*** | AIC | −4.8387 | $\alpha_2$ | 0.7539*** | AIC | -8.6462 |
| | (410.62) | | | | (27.7589) | | |
| Obs | 917 | SIC | −4.8020 | Obs | 917 | SIC | -8.6252 |
| Log-likelihood | 2232.808 | HQ | −4.8247 | Log-likelihood | 3141.189 | HQ | -8.6382 |

**Notes**: (1) ***, **, * represent statistical significance at 1%, 5%, and 10% levels, respectively. The z-statistics are presented in the brackets. (2) Due to the high non-linearity of the likelihood function, we use the starting values of parameters estimated from the simple GARCH (1,1) model.

average changes in the number of infected cases are around 1.14%, while the highest growth rate is approximately 294%. To our great surprise, the movement of equity market uncertainties is likely to be dominant with the widest spread of 423% changes. Among the proxies for market shocks, this factor experiences the most significant deviation with the highest value in terms of Standard deviation.

Table 3 reports the correlations between any two variables considered in our empirical analysis. The Panel A, B and C of Table 3 show the results about the whole sample, subsample with ending date January 20, 2020 and subsample with starting date January 21, 2020 respectively. The Panel B of Table 3 reports the correlation of *rEPU* and *Skew* is not significant with value of −0.0511; while the correlation of *rEPU* and *Skew* is significant at least 5% level with value of −0.1975 reported in the Panel C of Table 3. Moreover, the similar results about *rEMU* and *rEMV_ID* can be obtained by comparing the Panel B and Panel C of Table 3. The results of correlation provide an intuitive reflection that the economic policy uncertainty does not affect the stock market crash risk before the break of COVID-19, while the economic policy uncertainty have an impact on the stock market crash risk since the break of COVID-19.

### 3.5. Model specifications

We use the simple time series model to conduct our empirical analysis. The dependent variable is the stock market crash risk, measured by conditional skewness estimated from the GARCH-S model. We have two main explanatory variables. The first is the logarithmic growth rate of daily confirmed cases, *rCases*, and the logarithmic change rate of the Economic Policy Uncertainty Index for the United States, *rEPU*. The alternative variables for *rEPU* will be used in the robustness checks. The primary model is specified as:

$$Skew_t = c + \alpha \cdot Skew_{t-1} + \sum_{i=1}^{p} \beta_i \cdot rCases_{t-i} + \sum_{j=0}^{q} \lambda_j \cdot rEPU_{t-j} + \varepsilon_t \qquad (2)$$

**Table 3**
Correlation matrix of the related variables

| Panel A | The whole sample | | | | |
|---|---|---|---|---|---|
| **Variables** | *Skew* | *rCases* | *rEPU* | *rEMU* | *rEMV_ID* |
| *Skew* | 1.0000 | | | | |
| *rCases* | −0.0221 | 1.0000 | | | |
| *rEPU* | −0.0690** | 0.0068 | 1.0000 | | |
| *rEMU* | −0.0664** | 0.0151 | 0.2712*** | 1.0000 | |
| *rEMV_ID* | −0.0571* | 0.0049 | 0.0281 | 0.0253 | 1.0000 |
| **Panel B** | Subsample with ending date January 20, 2020 | | | | |
| *Skew* | 1.0000*** | | | | |
| *rCases* | N/A | N/A | | | |
| *rEPU* | −0.0511 | N/A | 1.0000*** | | |
| *rEMU* | −0.0545 | N/A | 0.2750*** | 1.0000*** | |
| *rEMV_ID* | −0.0133 | N/A | −0.0224 | −0.0342 | 1.0000*** |
| **Panel C** | Subsample with starting date January 21, 2020 | | | | |
| *Skew* | 1.0000*** | | | | |
| *rCases* | −0.0390 | 1.0000*** | | | |
| *rEPU* | −0.1975** | 0.0237 | 1.0000*** | | |
| *rEMU* | −0.1514* | 0.0543 | 0.2236*** | 1.0000*** | |
| *rEMV_ID* | −0.2102*** | 0.0070 | 0.4064*** | 0.4481*** | 1.0000*** |

**Notes:** ***, **, * represent statistical significance at 1%, 5%, and 10% levels respectively. The observations for Panel A are 917; The observations for Panel B are 761; The observations for Panel C are 156.

where $p$ and $q$ will be determined by the AIC or SC information criterion. In which, the c represents the constant term while $\alpha, \beta, \lambda$ are the coefficients of the Skew, cases, and EPU, respectively. We also look at the previous period with the time-lagged ($t-1$) to see the predictive power with the error terms ($\varepsilon_t$).

In order to examine the potentially different roles that EPU may play during the pandemic, we add the interaction term of *D_epid* and *rEPU* in our model:

$$Skew_t = c + d \cdot D\_epid + \alpha \cdot Skew_{t-1} + \sum_{i=1}^{p} \beta_i \cdot rCases_{t-i} + \sum_{j=0}^{q} \lambda_j \cdot rEPU_{t-j} + \sum_{j=0}^{q} \theta_j \cdot D\_epid \cdot rEPU_{t-j} + \varepsilon_t \quad (3)$$

The denotations in Equation (3) are similar to what we defined previously. However, instead of using one proxy to predict the market crashes, we also tried out to add the interaction terms, including the EPU and the period after the pandemic.

## 4. Empirical results

### 4.1. COVID-19 and stock market crash risk

We start our empirical analysis by only considering the explanatory variable *rCases*, and the results are shown in Table 4. As we expected, all the coefficient of *rCases*$_{t-1}$ is negative and significant at 1% level, indicating the severity of the pandemic has a direct negative impact on the crash risk of the US stock market. This finding is also consistent with Liu et al.(2020)'s work in the Chinese stock market. Furthermore, we also witness that the effects of the number of confirmed are not persistent because it only happens in one previous day after controlling the other previous days.

**Table 4**
The effects of COVID-19 on stock market crash risk

| Variables | (1) | (2) | (3) |
|---|---|---|---|
| *Intercept* | 0.0007 | 0.0010 | 0.0010 |
| | (0.4018) | (0.5655) | (0.5297) |
| *Skew*$_{(t-1)}$ | 0.1982*** | 0.1909*** | 0.1918*** |
| | (6.1851) | (5.8741) | (5.8797) |
| *rCases*$_{(t-1)}$ | −0.0245*** | −0.0340*** | −0.0345*** |
| | (-3.0197) | (-3.4662) | (-3.4816) |
| *rCases*$_{(t-2)}$ | | −0.0170* | −0.0151 |
| | | (−1.7227) | (−1.3789) |
| *rCases*$_{(t-3)}$ | | | 0.0037 |
| | | | (0.3760) |
| N | 919 | 918 | 917 |
| R² | 0.0500 | 0.0528 | 0.0529 |
| Adj-R² | 0.0479 | 0.0497 | 0.0488 |
| AIC | −2.9505 | −2.9505 | −2.9474 |
| SC | −2.9347 | −2.9295 | −2.9211 |

**Notes:** ***, **, * represent statistical significance at 1%, 5%, and 10% levels, respectively. The t-statistics are presented in the brackets.

While the number of cases is used as the good predictive factor to market changes on the onset of COVID-19 pandemic (Zhang et al. 2020; Ashraf 2020a; Ashraf 2020b), our findings are the first evidence to contribute to the extant literature that this factor could predict the market crashes. However, Goodell (2020) might argue that other factors could drive the equity market shocks. One of the potential determinants is the uncertainties which were raised from the economic and policy responses. In doing so, we used the EPU suggested by Baker et al. (2016) and Baker et al. (2020) to explore the market situation in this difficult time.

**4.2. Does and how does EPU matter?**

We further consider the role of economic policy uncertainty in affecting the crash risk in this section. We first only add the explanatory variable *rEPU* into our model, and the results are reported in Table 5. We can see that all the coefficients of $rEPU_t$ except that in column (4) is negative and significant at 5% level. It is consistent with our exception that increased economic policy uncertainty will increase the possibility of a stock market crash. Although the existence of economic and policy uncertainties exhibit the negative impacts on the probability of market downturn, these effects are less pronounced when adding up with the different time horizons. This implies that the weakly predictive power of Baker et al. (2020)'s proxy on the US market crash. Previously, the number of infected cases are likely to be robust when having different time zones while this magnitude of EPU is decreasing and turns into insignificant coefficients (in column 4). To our great surprise, we observed the strengthened effects of skewness, itself, by the increase in its coefficients (from 0.2032 to 0.2042, and significant at 1% level). The existing literature of Altig et al. (2020) offers insights about the different regimes in terms of EPU. In addition, several empirical evidence confirms the relationship between EPU and market volatilities by wave-let approaches (Choi 2020; Sharif et al. 2020) while our estimates shed new light on this linear dependence between market skewness and the standard proxy for the shaken markets. Notwithstanding its blur effects, in other words, improving the stability of economic policies will help prevent a stock market crash.

**Table 5**
The effects of EPU on stock market crash risk

| **Variables** | **(1)** | **(2)** | **(3)** | **(4)** |
|---|---|---|---|---|
| *Intercept* | 0.0004 | 0.0004 | 0.0004 | 0.0004 |
|  | (0.2392) | (0.2380) | (0.2387) | (0.2343) |
| *Skew$_{(t-1)}$* | 0.2032*** | 0.2037*** | 0.2037*** | 0.2042*** |
|  | (6.2845) | (6.2869) | (6.2830) | (6.3040) |
| *rEPU$_{(t)}$* | −0.0089** | −0.0084** | −0.0087** | −0.0070 |
|  | (−2.3390) | (−1.9999) | (-2.0153) | (−1.5786) |
| *rEPU$_{(t-1)}$* |  | 0.0012 | 0.0005 | 0.0030 |
|  |  | (0.2821) | (0.1060) | (0.5939) |
| *rEPU$_{(t-2)}$* |  |  | −0.0013 | 0.0027 |
|  |  |  | (−0.3035) | (0.5414) |
| *rEPU$_{(t-3)}$* |  |  |  | 0.0070 |
|  |  |  |  | (1.5728) |
| N | 917 | 917 | 917 | 917 |
| $R^2$ | 0.0460 | 0.0461 | 0.0462 | 0.0487 |
| Adj-$R^2$ | 0.0438 | 0.0429 | 0.0420 | 0.0435 |
| AIC | −2.9444 | −2.9423 | −2.9402 | −2.9407 |
| SC | −2.9286 | −2.9213 | −2.9139 | −2.9092 |

**Notes:** ***, **, * represent statistical significance at 1%, 5%, and 10% levels, respectively. The t-statistics are presented in the brackets.

When considering the influence of *rCases* and *rEPU* at the same time, we use the Akaike information criterion (AIC) and Schwarz criterion (SC) to determine the lag length of the variables. The results are shown in Table 6. Although the optimal lag lengths obtained according to different criteria are different, the conclusion that the severity of COVID-19 and EPU have negative influences on the crash risk of stock market is still holding. In terms of explanatory power, when the variables *rCases* and *rEPU* are added into the model simultaneously, its explanatory power has been significantly improved. For example, comparing the results in the first column of Table 6 with the results in the second column of Table 5, we can see that after adding *rEPU*, the *adj-R²* increases from 0.0497 to 0.0504. While comparing the results from the second column of Table 6 with that in the first column of Table 5, the *Adj-R²* has increased from 0.0479 to 0.0502. To sum up, our models remain robust when using the simultaneous appearance of both EPU and the number of infected cases. It is worth noting that both factors deteriorate the market by imposing the higher crashes on the onset of the pandemic outbreaks. However, we also have the caveat that these following estimates did not include the effects of the exact time of the pandemic. Schell et al. (2020) indicated that the adverse effects of COVID-19 are likely to be higher after the 21st January 2020, Public Health Risk Emergency of International Concern (PHEIC) announcements. In doing so, we will take a closer look at the market crash by adding the interactive terms to examine whether the EPU is strengthened after this event.

**Table 6**
The effects of COVID-19 and EPU on stock market crash risk

| **Variables** | (1) | (2) |
| --- | --- | --- |
| *Intercept* | 0.0010 | 0.0007 |
| | (0.5618) | (0.3952) |
| *Skew$_{(t-1)}$* | 0.1935*** | 0.2011*** |
| | (5.9587) | (6.2448) |
| *rCases$_{(t-1)}$* | −0.0335*** | −0.0241*** |
| | (−3.4234) | (-2.9812) |
| *rCases$_{(t-2)}$* | −0.0167* | |
| | (−1.7002) | |
| *rEPU$_{(t)}$* | −0.0086** | −0.0087** |
| | (−2.2788) | (−2.2954) |
| N | 917 | 917 |
| $R^2$ | 0.0582 | 0.0552 |
| Adj-$R^2$ | 0.0540 | 0.0521 |
| AIC | −2.9529 | −2.9519 |
| SC | −2.9266 | −2.9309 |

Notes: ***, **, * represent statistical significance at 1%, 5%, and 10% levels, respectively. The t-statistics are presented in the brackets. The optimal lag length in column (1) is determined by the AIC criterion, while the SC criterion determines the optimal lag length in column (2).

**4.3. Infected cases, EPU and the difficult time**

Finally, we consider whether the impact of EPU on crash risk is different during the pandemic. We know that when a crisis event occurs, the government usually needs to formulate a series of policies to alleviate the negative impact of the crisis. Therefore, economic policy uncertainty during the COVID-19 crisis may play a different role from the normal period.

The relevant results are given in Table 7. We can see that the coefficients of the interaction terms are negative and significant at the 5% level, while the coefficients of *rEPU* are not significant anymore. More accurately, take the results from column (1) for example, the coefficient of *rEPU* during the before-pandemic period is −0.0062 with t-statistics −1.5967 (not significant at 10% level). While during the pandemic period, the coefficient of *rEPU* is -0.0408 with t-statistics = −2.8666. It indicates that in a regular period, changes in economic policy uncertainty do not significantly correlate with the risk of stock market crashes. Moreover, during the pandemic, the increase in uncertainty caused by COVID-19 will increase the stock market crash risk. The timely formulation of response policies can reduce this uncertainty, thereby helping to prevent the possibility of extreme risks in the stock market.

**Table 7**

The different role of EPU during the pandemic

| Variables | (1) | (2) |
|---|---|---|
| *Intercept* | 0.0009 | 0.0009 |
|  | (0.4612) | (0.4653) |
| *D_epid* | −0.0011 | 0.0007 |
|  | (−0.2232) | (0.1365) |
| *Skew$_{(t-1)}$* | 0.1999*** | 0.1926*** |
|  | (6.2173) | (5.9408) |
| *rCases$_{(t-1)}$* | −0.0227*** | −0.0324*** |
|  | (−2.7921) | (−3.2291) |
| *rCases$_{(t-2)}$* |  | −0.0165 |
|  |  | (−1.64) |
| *rEPU$_{(t)}$* | -0.0062 | −0.0062 |
|  | (−1.5967) | (−1.5901) |
| *D_epid* rEPU$_{(t)}$* | −0.0345** | −0.0340** |
|  | (−2.3417) | (−2.3105) |
| N | 917 | 917 |
| $R^2$ | 0.0609 | 0.0637 |
| Adj-$R^2$ | 0.0557 | 0.0575 |
| AIC | −2.9536 | −2.9544 |
| SC | −2.9220 | −2.9176 |

**Notes:** ***, **, * represent statistical significance at 1%, 5%, and 10% levels, respectively. The t-statistics are presented in the brackets. We set the lag length of each variable based on the previous results in Table 6.

We can draw some interesting conclusions from our findings with the predictive power of both factors, namely EPU and infected cases when confronted with the role of the policy announcement. First, our findings are also in line with the existing literature about the prominent time, which might induce market shocks (Schell 2020; Goodell and Huynh 2020). More noticeably, while EPU did not have any correlation with equity market crashes before the coronavirus announcement, its effect turns out the observatory factor to the probability of the US equity market meltdown in the late period. It emphasizes the critical role of declaring the pandemic situation, which could drive the market changes. Second, the number of confirmed cases are still significant when we tried out different estimates. Therefore, our findings also confirm the extant literature of the predictive characteristic of a number of confirmed cases (Phan et al. 2020; Haroon et al. 2020). Before offering policy implications, we offer some different robustness checks to ensure what our empirical results are not spurious.

**4.4. Robustness results from alternative EPU indices**

There are two alternative measures about EPU: *rEMU* and *rEMV-ID*, representing the equity market uncertainties for the normal time and the pandemic period. In this section, we use these two variables to redo our empirical analysis. The results are reported in Table 8 and Table 9, respectively. In general, the results are robust.

One of the main points from Table 9 is the significant coefficients of the number of confirmed cases (*rCases*) in one previous term, implying that the increase in cases predicts the higher likelihood of a market crash. Besides, the Economic Uncertainty Index negatively correlates with the market downturn when confronting with the post-event announcement of coronavirus pandemic. Ultimately, what we found before still holds true when replacing with Equity Market Volatility: Infectious Disease Tracker (*EMV-ID*). Notwithstanding the current empirical evidence employing this proxy (Bai et al. 2020; Bouri et al. 2020), our results indicate the effects of market uncertainties with various proxies on the market crashes. Therefore, our results are still robust for further policy implications and discussion.

**Table 8**
Robustness results from EMU

| Variables | (1) | (2) | (3) | (4) |
|---|---|---|---|---|
| *Intercept* | 0.0007 | 0.0010 | 0.0009 | 0.0009 |
| | (0.3979) | (0.5710) | (0.4628) | (0.4676) |
| *D_epid* | | | −0.0011 | 0.0009 |
| | | | (−0.2324) | (0.1803) |
| *Skew$_{(t-1)}$* | 0.1968*** | 0.1890*** | 0.1994*** | 0.1912*** |
| | (6.1082) | (5.8143) | (6.1919) | (5.8901) |
| *rCases$_{(t-1)}$* | −0.0245*** | −0.0343*** | −0.0245*** | −0.0355*** |
| | (−3.0279) | (−3.4989) | (−3.0096) | (−3.5455) |
| *rCases$_{(t-2)}$* | | −0.0174* | | −0.0190* |
| | | (−1.7652) | | (−1.8824) |
| *rEMU$_{(t)}$* | −0.0034* | −0.0035* | −0.0023 | −0.0023 |
| | (−1.8934) | (−1.9330) | (−1.2211) | (−1.2328) |
| *D_epid* rEMU$_{(t)}$* | | | −0.0146** | −0.0153** |
| | | | (−2.1462) | (−2.2533) |
| N | 917 | 917 | 917 | 917 |
| $R^2$ | 0.0534 | 0.0567 | 0.0583 | 0.0619 |
| Adj-$R^2$ | 0.0503 | 0.0525 | 0.0531 | 0.0557 |
| AIC | −2.9501 | −2.9513 | −2.9508 | −2.9525 |
| SC | −2.9290 | −2.9250 | −2.9193 | −2.9157 |

**Notes:** ***, **, * represent statistical significance at 1%, 5%, and 10% levels, respectively. The t-statistics are presented in the brackets.

## 5. Conclusions

This paper has examined the role of economic policy uncertainty in affecting the stock market crash. we use the conditional skewness from the GARCH-S model as the proxy for crash risk. Our findings shed new light that COVID-19 will increase the crash risk of the US stock market. Second, we find that the increase in economic policy uncertainty has the risk of triggering a stock market crash. However, further evidence suggests that this effect of EPU is significant only during the pandemic. It reminds us that proactive policies should be implemented in time during crisis events to reduce the economic policy uncertainty, which is very important to prevent the stock market from crashing effectively.

There are some policy implications which could be raised from these following findings and results. First, not only investors but also policymakers should pay necessary attention to the growing of infected cases. Our findings emphasize the predictive power of this factor to the market shocks because the rising cases are

**Table 9**
Robustness results from EMV_ID

| Variables | (1) | (2) | (3) | (4) |
|---|---|---|---|---|
| Intercept | 0.0007 | 0.0011 | 0.0009 | 0.0009 |
| | (0.4020) | (0.5735) | (0.4625) | (0.4672) |
| D_epid | | | −0.0009 | 0.0011 |
| | | | (−0.1838) | (0.2189) |
| $Skew_{(t-1)}$ | 0.1987*** | 0.1909*** | 0.2000*** | 0.1919*** |
| | (6.1635) | (5.8727) | (6.2186) | (5.9194) |
| $rCases_{(t-1)}$ | −0.0236*** | −0.0333*** | −0.0208** | −0.0314*** |
| | (−2.9120) | (−3.3943) | (−2.5268) | (−3.1276) |
| $rCases_{(t-2)}$ | | −0.0172* | | −0.0185* |
| | | (−1.7492) | | (−1.8390) |
| $rEMV\_ID_{(t)}$ | −0.0063 | −0.0064 | −0.0014 | -0.0014 |
| | (−1.5778) | (1.6081) | (−0.3116) | (−0.3139) |
| $D\_epid* rEMV\_ID_{(t)}$ | | | −0.0258** | −0.0265*** |
| | | | (−2.5437) | (−2.6126) |
| N | 917 | 917 | 917 | 917 |
| $R^2$ | 0.0523 | 0.0555 | 0.0590 | 0.0625 |
| Adj-$R^2$ | 0.0492 | 0.0513 | 0.0539 | 0.0564 |
| AIC | −2.9489 | −2.9500 | −2.9516 | −2.9532 |
| SC | −2.9278 | −2.9237 | −2.9201 | −2.9164 |

**Notes:** ***, **, * represent statistical significance at 1%, 5%, and 10% levels, respectively. The t-statistics are presented in the brackets.

considered as the signal to the market movement to the negative side. Thus, the tracking record is mandatory to avoid the extreme event, which might cause huge losses when investing in this challenging time. Second, the role of news, mainly since January with restrictions of travelling, national curfews, the bans on public gatherings, and social distancing, would induce the harmful shakes of the market. Thus, it is an underlying mechanism to transmit the potential risk to the market crash after controlling other factors. Concomitantly, the process of reviewing the news content might help investors avoiding the sudden adverse shocks when the frequencies of negative words, relevant to the uncertainties because these elements are contributing to the market uncertainty and then its correlation with the market meltdown.

A caveat should be included in this analysis, that the COVID-19 pandemic is still in a nascent period of rapid growth. Thus, what may be right for the situation currently is likely to change until the other outbreak, the appearance of vaccine, or other factors, in some manner. Hence, this study just provides the preliminary results to understand how the infected cases and EPU are correlated with the market crash. Future research might focus on the dynamics of this effect. Thus, this avenue is still promising indeed.